\UseRawInputEncoding
\def\preprint{1}			

\ifdefined\preprint
    \documentclass[preprint, 12pt]{elsarticle}
\fi
\ifdefined\wordcount
  \documentclass[final,3p,times,twocolumn]{elsarticle}
\fi
\ifdefined\final
  \documentclass[final,3p,times,twocolumn]{elsarticle}
\fi
\usepackage{tcolorbox}
\usepackage{amssymb}
\usepackage{booktabs}
\usepackage[version=4]{mhchem}
\usepackage{siunitx}
\usepackage{pgfplots}
\usepackage{xspace}
\usepackage{pgf-umlcd}

\usepackage{tikz}
\usepackage{framed} 
\usepackage{multicol} 
 
\usepackage{nomencl} 
\makenomenclature
\renewcommand*\nompreamble{\begin{multicols}{2}}
\renewcommand*\nompostamble{\end{multicols}}
\pgfplotsset{compat=newest}
\usepgfplotslibrary{groupplots}
\usepgfplotslibrary{dateplot}
\usetikzlibrary{shapes,arrows}
\usetikzlibrary{matrix, positioning, fit}
\tikzstyle{block} = [rectangle, draw, fill=black!20, 
text width=10em, text centered, rounded corners, minimum height=1em]
\tikzstyle{line} = [draw, -latex']
\tikzstyle{cloud} = [draw, ellipse,fill=green!20, node distance=1.65cm,
minimum height=1em]

\usepackage{lineno}
\usepackage{hyperref}
\hypersetup{
    colorlinks=true,
    linkcolor=blue,
    filecolor=magenta,      
    urlcolor=cyan,
}
\biboptions{sort&compress}

\usepackage{subfig}
\usepackage{algorithm}
\usepackage{algpseudocode}%
\usepackage[export]{adjustbox}
\usepackage{float}
\usepackage{multirow}
\usepackage{graphicx}
\definecolor{darkgreen}{rgb}{0.0, 0.5, 0.0}
\newcommand{\nproc}{\textit{N\textsubscript{p}}\xspace}
\newcommand{\nprob}{\textit{N\textsubscript{c}}\xspace}
\newcommand{\nsend}{\textit{N\textsubscript{s}}\xspace}

\usepackage{tabto}

\restylefloat{table}

\begin{document}

\begin{frontmatter}



\title{DLBFoam: An open-source dynamic load balancing model for fast reacting flow simulations in OpenFOAM}


\author[fir]{Bulut Tekg\"{u}l\corref{cor1}}
\ead{bulut.tekgul@aalto.fi}
\author[fir]{Petteri Peltonen}
\author[sec]{Heikki Kahila}
\author[fir]{Ossi Kaario}
\author[fir]{Ville Vuorinen}

\address[fir]{Department of Mechanical Engineering, Aalto University School of Engineering, Otakaari 4, 02150 Espoo, Finland}

\address[sec]{W\"{a}rtsil\"{a} Finland Oy, 65101 Vaasa, Finland}

\cortext[cor1]{Corresponding author: Bulut Tekg{\"u}l}

\begin{abstract}
{Computational load imbalance is a well-known performance issue in multiprocessor reacting flow simulations utilizing directly integrated chemical kinetics.} We introduce an open-source dynamic load balancing model {named \texttt{DLBFoam}} to address this issue within OpenFOAM, an open-source C++ library for Computational Fluid Dynamics (CFD). Due to the commonly applied operator splitting practice in reactive flow solvers, chemistry can be treated as an independent stiff ordinary differential equation (ODE) system within each computational cell. As a result of the highly non-linear characteristics of chemical kinetics, a large variation in the convergence rates of the ODE integrator may occur, leading to a high load imbalance across multiprocessor configurations. However, the independent nature of  chemistry ODE systems leads to a problem that can be parallelized easily ({called an} \textit{embarrassingly parallel} {problem} in the literature) during the flow solution. The presented model takes advantage of this feature and balances the chemistry load across available resources. Additionally, a reference mapping model is utilized to further speed-up the simulations. When \texttt{DLBFoam} it utilized with both these features enabled, a speed-up by a factor of 10 is reported for {reactive flow benchmark cases.} To the best of our knowledge, this model is the first open-source implementation of chemistry load balancing in the literature.

\end{abstract}


\end{frontmatter}

{\bf PROGRAM SUMMARY}

\begin{small}
\noindent
{\em Program Title: }   DLBFoam                                   \\
{\em CPC Library link to program files:} (to be added by Technical Editor) \\
{\em Developer's repository link:} \url{https://github.com/blttkgl/DLBFoam} \\
{\em Code Ocean capsule:} (to be added by Technical Editor)\\
{\em Licensing provisions:} GPLv3 \\
{\em Programming language: } C++                   \\
{\em Supplementary material:}                                 \\
{\em Nature of problem:} Solution of chemical kinetics in parallel reacting flow solvers raises a computational imbalance across multiprocessor architectures. DLBFoam balances the load distribution evenly, providing significant speed-up in reacting CFD applications.\\
{\em Solution method:} The dynamic load balancing is implemented by distributing the point-wise chemistry problems from most loaded processes to less loaded ones using MPI communication protocol. \\
{\em Additional comments including restrictions and unusual features:} The present model is designed to work with the standard chemistry model class available in OpenFOAM (versions 7 and 8). For the time being, the model does not support derived combustion models such as ''TDAC'' and covers gas-phase reaction kinetics only. In addition, the boundary surface chemistry problems are neglected by the model.
\end{small}

\section{Introduction}
\label{section::introduction}

The development of efficient Computational Fluid Dynamics (CFD) simulation tools for reacting flows is a crucial step in the research and development of less polluting combustion concepts \cite{Poinsot2001}. As the need for more detailed combustion models has become prominent, chemical kinetics models have grown in size and complexity, resulting in higher computational cost and often exceeding the computational cost of fluid dynamics by a factor of 100 \cite{Peters2000}. Even though most CFD software utilize distributed-memory parallel architectures to their full extent, incorporating an efficient solution for reacting flows tends to lead to a high computational load imbalance during  parallel execution \cite{Muela2019}.

Commonly, engineering CFD codes assume an operator-splitting strategy in the reacting flow solver implementation, enabling the decoupling of the chemistry and fluid dynamics in the reacting flow solution during the calculation of the chemical source terms in the governing equations \cite{Law2006,Speth2013}. {Therefore, the changes in thermochemical composition $\Phi$ (temperature, pressure, and species concentrations) due to chemical reactions per computational element (e.g.~finite-volume cell) are computed by solving a cell-wise independent homogeneous reaction system. The reaction system describes the rate of change of thermochemical composition as a stiff system of ordinary differential equations (ODEs), i.e.~$\partial\Phi / \partial t = f(\Phi, t)$ \cite{Law2006}.}

\begin{figure}[H]
\centering
{\includegraphics[width=0.47\columnwidth]{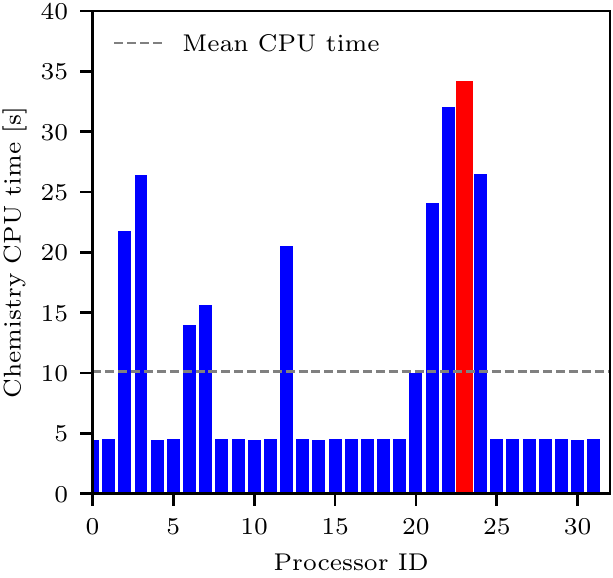}}
\caption{An illustration of the computational imbalance in a reactive CFD simulation. The cost of the chemistry solution across different processes within a given CFD time step varies, creating a bottleneck at the process with highest computational load (marked with red). {Data, plotting scripts and figure are available under the CC-BY license \cite{data}.}}
\label{fig:imbalance_illustration}
\end{figure}

Typically, the computational cost of the chemical source term evaluation dominates the performance metrics and creates an uneven computational load distribution in parallel applications. The difficulties occur due to the intrinsic and non-linear nature of the ODE system. As the computational cost of solving the associated stiff ODEs scales quadratically with the number of species \cite{Niemeyer2016}, detailed reaction mechanisms easily become impractical. Furthermore, due to the vast scale separation between the fastest and slowest chemical reaction time scales, the system of ODEs is practically always numerically stiff, requiring the use of implicit time integrators with low time step values \cite{Hairer1996}. The time step value and hence the total number of floating-point operations during the integration depends on the initial thermochemical composition \cite{Stone2018}. Furthermore, in most CFD codes parallelization is achieved with geometrical domain decomposition, leading to explicit chemistry load imbalance due to spatially and temporally varying $\Phi$ values. A visual illustration of the chemistry load imbalance is presented in Figure~\ref{fig:imbalance_illustration}.

Apart from the optimization of the ODE solver algorithms and implementations \cite{Niemeyer2016, Curtis2019, Imren2016}, several modeling strategies have been proposed to reduce the high computational requirements associated with reacting flows. To name a few successful approaches: in-situ adaptive tabulation method \cite{pope_isat}, dynamic and adaptive reduction methods for chemical kinetics \cite{Ren2014a} and dynamic stiffness removal methods \cite{Lu2009} each provide performance gains to a certain extent. However, these strategies are often formulated in a processor-based approach, being still prone to yield computational load imbalance across available resources \cite{Contino2011a}. 

In the context of the present study, dynamic load balancing of chemistry has been previously covered in literature. One method to mitigate the load imbalance is to use a custom decomposition with a prior knowledge on the spatial activity of chemical kinetics \cite{Turkeri2019}. However, in large complex geometries, custom decomposition becomes impractical \cite{Turkeri2019,Cuoci2013}. In terms of dynamic run-time load-balancing algorithms, Antonelli et al.~\cite{Antonelli2011} developed a Message-Passing Interface (MPI) based parallel solver which utilizes a cell distribution based load balancing algorithm. Both Kodavasal et al.~\cite{Kodavasal2016} and Shi et al.~\cite{Shi2012} considered stiffness detection approaches as balancing criteria for their balancing algorithms. Recently, Muela et al.~\cite{Muela2019} presented a dynamic load balancing method which also utilized a stiff cell detector to choose the optimal ODE integration method. In practical benchmark cases, the aforementioned methods reported speed-up factors in the range three to five.

In contrast to previous studies where implementations are either not publicly available or based on commercial CFD codes, in the present study we introduce a robust open-source load balancing algorithm for parallel reacting flow simulations. To the best of our knowledge, this model is the first open-source implementation of chemistry load balancing in literature. In addition to the load balancing model, a zonal reference mapping model is also introduced to further improve the available balancing performance. The implementation is carried out in OpenFOAM, an open-source C++ library targeted for CFD applications \cite{Weller1998}. It is worth noting that an earlier version of this now publicly shared model has already been utilized in multiple published combustion studies by the authors \cite{Kahila2019,Kahila2019a,Tekgul2020}.

The paper is outlined as follows: The implemented dynamic load balancing algorithm is presented  in Section \ref{section:algorithm}. Section \ref{section:benchmark} provides benchmarks highlighting the theoretical performance gains of this model, while Section \ref{section:Results} reports performance metrics in practical reacting flow configurations.  Section \ref{section:conclusions} concludes the study with a brief discussion on the results along with the potential for further improvements.

\section{\label{section:algorithm} Implementation details}

\subsection{Reacting flow solver}

\begin{figure}[H]
\centering
\resizebox{0.6\columnwidth}{!}{%
	\begin{tikzpicture}[font=\small,
	mymatrix/.style={matrix of nodes, nodes=typetag, row sep=0em},
	mycontainer/.style={draw=blue, inner sep=1ex},
	typetag/.style={draw=gray, inner sep=1ex, anchor=west},
	title/.style={draw=none, color=black, inner sep=0pt}
	]
\node [block,fill=none] (rhoEqn) {{Solve mass conservation}};
\node [block,fill=none, below of=rhoEqn] (UEqn) {Solve momentum};
	\matrix[mymatrix,below of=UEqn,yshift=-2.5cm] (mx1) {
	|[title]|Solve chemistry (\texttt{DLBFoam})\\
	\node [block,fill=lightgray] (refMap_m) {Stage 1: Pre-processing};
	\node [block,fill=lightgray,below of=refMap_m] (balance_m) {Stage 2: Balancing};
	\node [block,fill=lightgray,below of=balance_m] (solve_m) {Stage 3: Solution};
	\node [block,fill=lightgray,below of=solve_m] (unbalance_m) {Stage 4: Update};\\
};
\node[mycontainer, fit=(mx1)] (Chemistry) {};
\node [block,fill=none, below of=Chemistry,yshift=-2.5cm] (Transport) {Solve species transport};
\node [block,fill=none, below of=Transport] (EEqn) {Solve energy};
\node [block,fill=none, below of=EEqn] (pEqn) {Solve pressure};
\node [left of=Chemistry,fill=none, xshift=-3cm, align=left] (PIMPLE) {PIMPLE \\ iterations};
\draw [->] (rhoEqn.south) -- (UEqn.north);
\draw [->] (UEqn.south) -- (Chemistry.north);
\draw [line] (EEqn) -- (pEqn);
\draw [line] (Chemistry) -- (Transport);
\draw [line] (Transport.south) -- (EEqn.north);
\draw[->] (pEqn.west) -| (PIMPLE.south);
\draw[->] (PIMPLE.north) |- (rhoEqn.west);	
	\end{tikzpicture}}
\caption{A schematic showing the main steps of the reactive solver within a single CFD time step. The \texttt{DLBFoam} part is highlighted in the figure.}
\label{fig:cfd_schematic}
\end{figure}
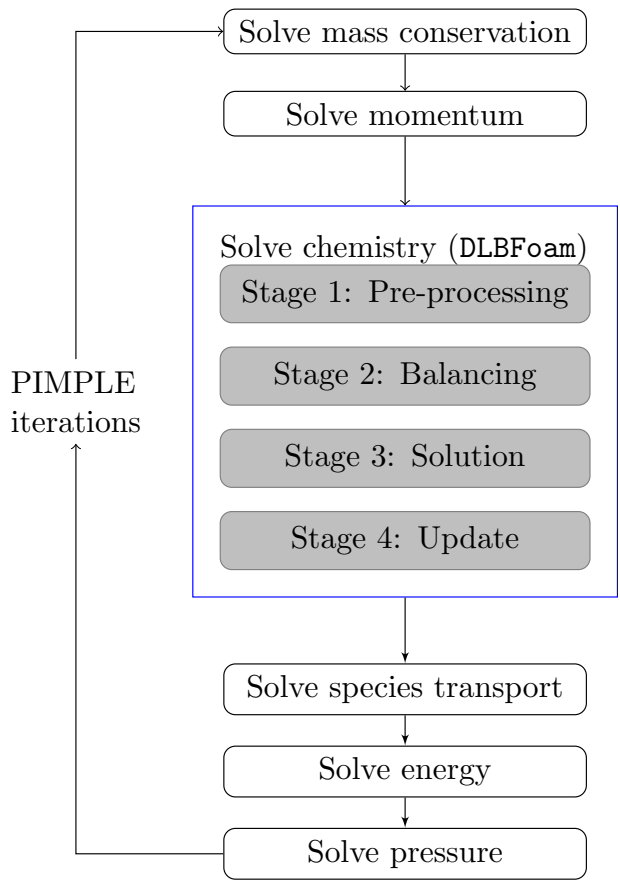

A schematic of the different stages of the reacting flow solver used in the paper is presented in Figure~\ref{fig:cfd_schematic}. Solution for mass, momentum, species, and energy conservation laws is achieved in an iterative manner. The \texttt{reactingFoam} and \texttt{sprayFoam} reactive solvers of OpenFOAM~\cite{Weller1998}  with compressible PIMPLE algorithm are used. Within each time step, a Poisson equation for pressure is solved and the velocities obtained from the solution of the momentum equation are corrected to ensure that the mass is conserved. 

Here we consider the solution of chemistry source terms (i.e.~net production/consumption rates of species concentrations) as a separate step, as highlighted in Fig.~\ref{fig:cfd_schematic}. In the present study, the chemistry source terms are evaluated by direct integration of chemical kinetics with no additional models. The source term evaluation is performed in a separate module which is interfaced at the solver level (chemistry model). The implementation of the load balancing algorithm is carried out in this module as a separate compilation unit. The details of the algorithm are discussed in the following subsection.

\subsection{Dynamic load balancing}
\label{subsec:balancing}
As mentioned before, the reaction source term for an individual CFD cell can be computed independently from all other cells. However, due to the non-linear nature of chemistry, the floating point operations required to evaluate source terms can vary significantly from cell to cell. {In most CFD simulations, the computational geometry is decomposed using some of the available domain decomposition algorithms. However, dividing the domain geometrically into chunks of close to equal cell count may lead to chemical imbalance due to the different convergence rates of chemistry ODE problems within processes, as illustrated in Fig.~\ref{fig:imbalance_illustration}. Hence, it is beneficial to send some of the chemical problems from the overloaded processors (senders) to the less occupied processors (receivers) for more uniform load balancing. Obviously, once the chemistry calculation has been completed, the roles interchange once the updated chemical solutions are sent back to where they  physically belong i.e. as source terms in the species and enthalpy equations.}

\begin{figure}[H]
\centering
\includegraphics[width=67mm]{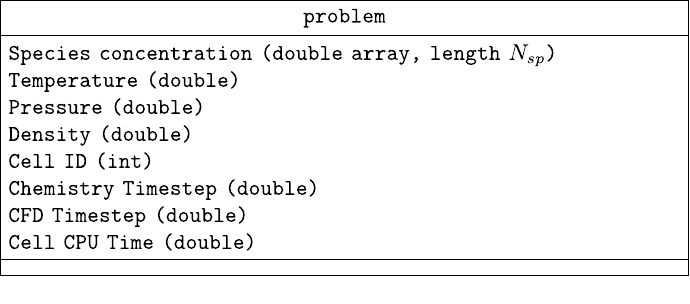}
\caption{{Data structure of a \texttt{problem} that contains the necessary information to solve the chemistry ODE system of the cell \textit{n}. In C++, the \texttt{problem} is implemented as a \texttt{struct} data type. $N_{sp}$ denotes the number of species in the utilized chemical kinetics mechanism.}}
\label{fig:problemschematic}
\end{figure}

The dynamic load balancing algorithm implemented in the present paper consists of four stages: 1) preprocessing stage, 2) balancing stage 3) solution stage, and 4) update stage. In the preprocessing stage, the thermochemical variables and other variables needed for the ODE solution for each cell $n$ are packed into a single data structure, here denoted as a \texttt{problem}, and an array of these structures is formed. {Figure \ref{fig:problemschematic} describes the variables within a \texttt{problem} data structure needed to describe a single chemistry ODE problem. A \texttt{problem} consists of $N_{sp}$+6 double precision floating point variables and one integer variable. Here, $N_{sp}$ is the number of species in the utilized chemical kinetics mechanism. The amount of data to be transferred between a sender-receiver pair depends on various case setup parameters such as the size of the utilized chemical mechanism, number of cells per process and the level of imbalance between a sender-receiver pair. }

Next, stage 2 is discussed together with a pseudocode illustration  presented in Algorithm \ref{alg:lb_algorithm}. Stage 2 is initiated by defining a processor-based load value $l_k$ for all available \nproc processes (line 1). Here, we define the load $l_k$ as the sum of total CPU time spent on solving \nprob problems on process $k$. To compute the global balancing statistics, a sorted global list of loads {$L=\left[l_0, l_1,...,l_{\nproc-1}\right]$} is formed and broadcasted to all available processes (line 2). Next, the algorithm aims at setting the global arithmetic mean load ($\overline{L}$) to all processes by finding sender-receiver rank pairs $(r_s, r_r)$ sharing a load $\Delta L_{send}$ (line 8). Then, the CPU time based load value is converted to \nsend chemistry problems to be transferred between the sender-receiver pair (line 11). {It is important to note that the smallest amount of load that can be sent between processes is controlled by explicitly removing very small load balancing communications and not executing them. This is done by ensuring that for a given sender-receiver pair,  $\Delta L_{send} \geq 0.01 \times  \overline{L} $  is satisfied.} Finally, $r_s,r_r$ and \nsend are inserted into a list denoted as \texttt{operations} (line 12), which is utilized to execute multiple non-blocking MPI calls to communicate the problems across processes (line 21). {Note that in the case of a sender with multiple receivers or a receiver with multiple senders, the communication protocol is organized so that all the data are sent/received simultaneously.}


In stage 3, the distributed chemistry problems are solved by their new processes. We note that the input data (a problem) is tagged with a cell id and the tag is copied to the output data (a solution) to ensure the correct placement of the reaction rate in the CFD domain in the original process.  In stage 4, the solutions are communicated back to the original processes based on the same $(r_s, r_r, \nsend)$ information from stage 2 and the chemical source term values of the cells are updated. {It should be noted that the stage 4 also includes an explicit MPI barrier to ensure that the processes update the reaction rates only after all the problems are communicated back to their original processes.}

\renewcommand{\Comment}[2][.45\linewidth]{\leavevmode\hfill\makebox[#1][l]{//~#2}}
\begin{algorithm}[H]
	\begin{algorithmic}[1]
	    \State $l_k \gets$ $\sum\limits_{n=1}^{\nprob}t_{\mathrm{cpu}}^n$ \Comment {load for rank $k$ at given time}
	    \State $L \gets$ $\mathrm{allGather}(l_k)$ \Comment {List of loads $L=\left[l_0, l_1,...,l_{\nproc-1}\right]$}
	    \State $\overline{L} \gets \sum\limits_{i=0}^{\nproc-1} l_k/\nproc$ \Comment {average load over all processes}
	    \newline
        \State $L, R \gets \mathrm{sort}(L)$ \Comment{Sort $L$, and get a new rank indexing  $R$}
        \newline
        \State $i \gets 0$
        \State $j \gets {\nproc-1}$
		\While {$ i \neq j$}
        \State $\Delta L_{send} \gets \min(\overline{L} - L_i , L_j - \overline{L})$ \Comment {Define load to be transferred}
        \State $r_r \gets R_i$ \Comment {Rank ID receiving data}
        \State $r_s \gets R_j$ \Comment {Rank ID sending data}
        \State find \nsend, s.t. $\Delta L_{send} \approx \sum\limits_{n=1}^{\nsend}t_{\mathrm{cpu}}^n$ \Comment{Corresponding \nsend problems}
        \State \texttt{operations.insert}($r_s$, $r_r$, \nsend) \Comment{List of balancing operations}
        \newline
        \State $L_i \gets L_i + \Delta L_{send}$ \Comment{Increase by the send value}
        \State $L_j \gets L_j - \Delta L_{send}$ \Comment{Decrease by the send value}
        \If{$|L_i - \overline{L}| \approx 0$}
          \State $i \gets i + 1$ \Comment {Move to next receiver}
        \Else
          \State $j \gets j - 1$ \Comment {Move to next sender}
        \EndIf
        \EndWhile
        \newline
        \State \texttt{process(operations)} \Comment {Call MPI functions}
        \State \Comment {based on the operations list.}
		\caption{\label{alg:lb_algorithm} Pseudo-code representation of the load balancing algorithm stage two over \nproc processes w.r.t.~a $n$th cell CPU times $t_{\mathrm{cpu}}^n$ at a time $t=n+1$.}
	\end{algorithmic}
\end{algorithm}

\subsection{Reference mapping}

In addition to the load balancing, we have implemented a simple reference mapping feature which allows us to further reduce the computational cost. The aim is to group cells sharing similar {thermochemical composition ($\Phi$)} values together and solve the chemistry only once for this group. Such a mapping approach is intended to be used for regions with low reactivity, e.g.~where no fuel is present. {Reference mapping model is very similar to multi-zone reduction models found in commercial CFD software \cite{Mandhapati}. However, by design our approach is far more conservative since it is mostly intended for mapping non-reacting mixtures including no fuel to one another.} In our implementation, the reference mapping acts as a filter in stage 1 of the load balancing algorithm, where the reaction rates of cells satisfying {a user-given} criteria are copied from a reference cell solution. At a {given} time instance, a reference cell is picked and the chemistry source term of that cell is solved and copied to other reference cells. The criteria used for identifying the reference cells is:

\begin{equation*}
\begin{split}
&Z_i < Z_{tol}, \\
&|T_i - T_{ref}| < T_{tol},
\end{split}
\label{eq:reference_cell_criteria}
\end{equation*}

\noindent where $Z_i$ and $T_i$ are mixture fraction and temperature of $i$th cell, respectively, $T_{ref}$ denotes the temperature of the chosen reference cell and $Z_{tol}$ and $T_{tol}$ denote the user-given tolerance values. In the reacting flow community, the mixture fraction stands for a conserved scalar describing the mixing state of a fuel and oxidizer uniquely. Although there are different definitions of $Z$ in the literature, in this paper the Bilger's definition \cite{Bilger1990} is used. The reference solution is computed from the first reference cell found, and this solution is then mapped to the subsequent reference cells.

The reference mapping implemented in the present study is applied to each process separately. Depending on the value set of $Z_{tol}$ and $T_{tol}$, as well as whether the temperature criterion is applied or not, the mapped solution and the actual solution may be slightly different from one another. As long as proper tolerances are used, the introduced error is rather small and does not affect the global characteristics of the reactive simulation. According to the authors' experience, the error is negligible compared to typical uncertainties found in various combustion models. Details on the validation of the reference mapping against the standard model as well as a discussion on the level of error introduced are presented in \ref{appendixA}. {We emphasize that the reference mapping model improves the balancing performance significantly by further reducing the load of the more idle processes and increasing their potential to receive more load from the busier processes. However, due to its process-based formulation it often provides infinitesimal speedup if it is utilized without the load balancing. It is also worth noting that reference mapping may introduce some approximation to the chemistry solution especially at the decision boundary between the mapped and unmapped regions. Therefore, it should be utilized with caution like any other zonal reduction model.}

\section{Unit benchmarks}
\label{section:benchmark}

In this section, the performance of the load balancer model is demonstrated by using a simple test environment created in OpenFOAM. Inspired by the performance analysis presented by Muela et al. \cite{Muela2019}, we present a similar analysis to show the efficiency as well as the scaling performance of our load balancer model for: i) different number of processes and ii) different number of cells per process. 

To carry out the performance analysis, we have developed a benchmarking suite in OpenFOAM, where a given set of chemistry problems are solved in parallel by the ODE solver and the total execution times are measured. Two different \textit{pseudo-problems} are predefined: a \textit{heavy problem}, which is a stiff ODE problem and takes longer to compute, and a \textit{light problem}, which is a less stiff problem with a shorter solution time. Here, a benchmark environment is created by defining the number of chemistry problems (\nprob), available resources (\nproc processes), and a predefined $\theta$ value, describing the ratio of \textit{heavy problem}s to the number of problems in each process. 

Throughout the benchmark analysis, the ratio of all \textit{heavy problem}s to the total number of problems across all processes is fixed as 0.2. Different configurations are then created to distribute the total heavy load either evenly (balanced load configuration) or unevenly (unbalanced load configuration). The following benchmark configurations are investigated:
\begin{itemize}
    \item $C_1$: "very unbalanced", $\theta$ = 1 for 20\% of \nproc
    \item $C_2$: "slightly unbalanced", $\theta$ = 0.8 for 25\% of \nproc
    \item $C_3$: "slightly balanced", $\theta$ = 0.4 for 50\% of \nproc
    \item $C_4$: "very balanced", $\theta$ = 0.2 for 100\% of \nproc.
\end{itemize}

The benchmark statistics are obtained by running 10 separate samples for each configuration to minimize the variance due to any hardware or parallel communication issues. Each case is run first with the unbalanced (standard) model, then with the balanced model, and the speed-up ratio $\chi_{su}$ for each condition is calculated as:

\begin{equation}
    \centering
   \chi_{su} =  \frac{\tau_{unb}}{\tau_{bal}},
\end{equation}

where $\tau_{unb}$ and $\tau_{bal}$ are the mean execution times over all samples of unbalanced and balanced cases, respectively. {In addition, using the ratio of a \textit{heavy} problem to a \textit{light} problem ($\xi$), the maximum theoretical speed-up that can be attainable for cases C$_1$-C$_4$ are calculated as:}

\begin{equation}
     \frac{\theta\times\xi+(1-\theta)}{\frac{\theta\times\xi+(1/x-1)}{1/x}},
\end{equation}

\noindent{where $x$ is the percentage of processes which $\theta$ condition applies for each condition (for instance, it is 20\% (0.2) for C$_1$).}

All simulations in this paper are carried out by the Mahti supercomputer, provided by the CSC - Finnish IT Center for Science. Mahti contains 1404 nodes with two 64 core AMD EPYC (Rome) processors, each node running at 2.6 GHz, as well as high-speed 200 Gpbs infiniband HDR interconnection network between the nodes, all running on a RHEL 7.8 Linux operating system \cite{Mahti}.

\begin{figure}[H]
    \subfloat[\label{fig:benchmark_ncpu}]{{\includegraphics[width=0.48\columnwidth]{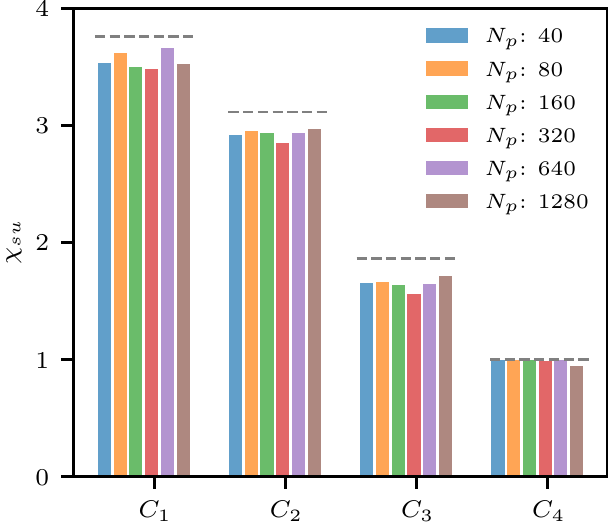} }}%
    \qquad
    \subfloat[\label{fig:benchmark_nprob}]{{\includegraphics[width=0.48\columnwidth]{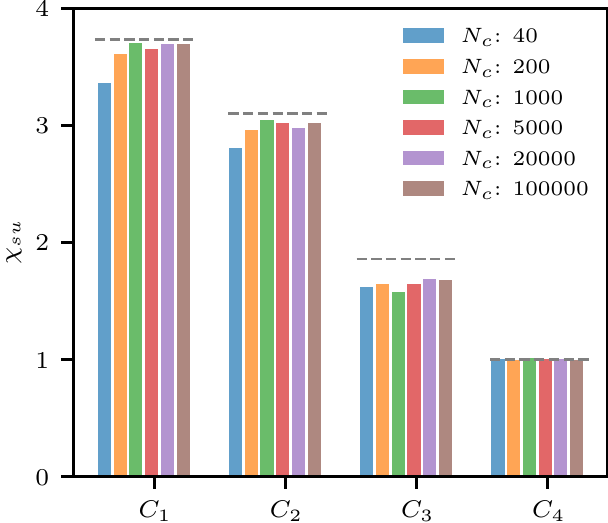} }}%
	\caption{{Speed-up ratio of a) different process counts \nproc with number of problems \nprob = 200  b) different \nprob with \nproc = 80, from imbalanced $(C_1)$ to balanced $(C_4)$ configurations. The dashed lines show the maximum theoretical speed-up that can be obtained for each configuration.} {Data, plotting scripts and figure are available under the CC-BY license \cite{data}.}}
	\label{fig:benchmark}
\end{figure}

Figure \ref{fig:benchmark} shows the benchmark results for configurations $C_1$-$C_4$ with varying \nproc and \nprob values. Figure \ref{fig:benchmark_ncpu} highlights the steady performance over a large range of process counts with a fixed \nprob = 200. It can be seen that the balancing algorithm scales up to \nproc = 1280 with no apparent performance issue. In addition, the $\chi_{su}$ value increases with increasing imbalance ($C_4 \rightarrow C_1$), since the potential balancing gain to be obtained is higher for cases with higher imbalance.

Figure \ref{fig:benchmark_nprob} shows the influence of problem count \nprob to balancing performance for a fixed \nproc = 80. It can be clearly seen from the high imbalance condition ($C_1$) that the $\chi_{su}$ first increases with increasing \nprob from 40 up to 1000, then it stays in the same level for higher \nprob values. The reason is two-fold: i) for lower \nprob values the global mean load value $(\overline{L})$ that the balancer tries to reach is smaller, therefore the cost of communication is comparable to the cost of gain obtained from the balancing, and ii) for lower \nprob values the number of problems describing the extra load is very sensitive, e.g. the balancer may overshoot/undershoot the target load condition with a greater margin by sending one cell more/less with lower \nprob values. Once the \nprob value is large enough so that the communication overhead is insignificant and the number of problems that are being transferred between processes are large enough to create the balancing statistics accurately, the $\chi_{su}$ is stabilized.

One common characteristic observed with both configurations presented in Figure \ref{fig:benchmark} is the change in balancing performance with respect to the level of imbalance in the benchmark. While the increase in $\chi_{su}$ with increased imbalance is already discussed, it is also important to note that $\chi_{su}\approx 1$ for the very balanced case ($C_4$). The $\chi_{su}$ values close to 1 indicate that the operations that create the load balancing statistics and determine the inter-processor communications are not resulting in significant overhead compared to ODE solution. Since the case is already perfectly balanced, the balancer does not communicate any data between the processes. {A more detailed analysis on the computational overhead associated with the dynamic load balancing algorithm is presented in \ref{appendixB}.}

In summary, this benchmarking analysis shows that the balancing algorithm described in Section \ref{subsec:balancing} shows very good performance and scalability for a range of process counts and the number of chemistry problems per process. The selected values for these two parameters are typical for reacting CFD simulations, ranging from small simulations that can be run on personal computers to very large simulations that require high-performance computing clusters.

\section{Results}
\label{section:Results}
Following the performance benchmark of the load balancing model in the previous section, here we demonstrate the model performance in actual reactive CFD simulations. First, a 2D reactive shear layer case with a uniform mesh is simulated to show the effects of load balancing, reference mapping model, number of processes, and decomposition methods. Furthermore, a 3D reactive spray LES simulation is performed to indicate a potential speed-up that can be obtained in large-scale reactive simulations. We do not present physics-based case validation, but aim to show the performance gain available with a load balancing algorithm.

\subsection{Reactive shear layer}
\label{subsec:shearlayer}
\begin{figure}[H]
 \subfloat[\hspace{1.6cm}]{{\includegraphics[height=5.6cm]{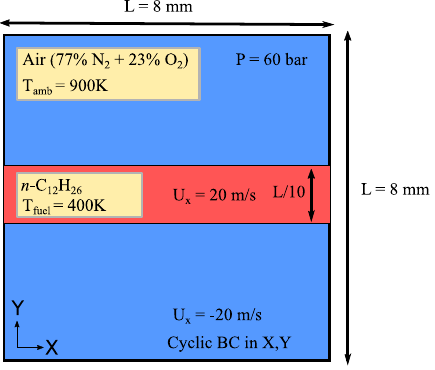} }}%
    \qquad
    \subfloat[\hspace{1.6cm}]{{\includegraphics[height=5.6cm]{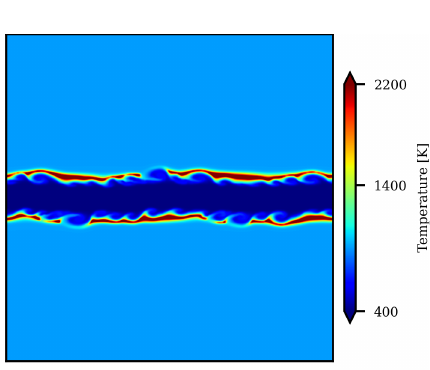} }}%
\caption{a) Schematic of the test case and b) an instantaneous representation of the temperature field during the simulation.}
	\label{fig:schematic}
\end{figure}

A schematic describing the 2D reactive shear layer configuration is presented in Figure \ref{fig:schematic}. A square domain ($8\si{\milli\meter}\times8$\si{\milli\meter}) with cyclic boundary conditions is discretized by a structured mesh with $400\times400$ resolution. A hyperbolic tangent function is utilized to generate a smooth shear layer of $L/10$ width in the domain. The momentum shear layer is characterized by a relative velocity difference of  $\Delta U_{x}$ = \SI{40}{\meter/\second}. The shear layer instability is initiated by introducing a sinusoidal perturbation to the vertical velocity component $U_y$. 

The fuel, $n$-dodecane ($n$-\ce{C12H26}), is placed in the middle while air (0.77 \ce{N2} + 0.23 \ce{O2}) is set elsewhere. The fuel and air temperatures are set to $T_{fuel}$ = \SI{400}{\kelvin} and  $T_{amb}$ = \SI{900}{\kelvin}, respectively. Constant pressure of $p$ = 60 bar is initialized in the domain. The chemical kinetics is modeled by the skeletal mechanism developed by Yao et al. \cite{Yao2017}, including 54 species and 269 reactions. The relative and absolute tolerances of the ODE solver tolerances are set to 1e$^{-5}$ and 1e$^{-8}$, respectively. The reference cell would be chosen at an arbitrary point outside the shear layer with $Z_{tol}$ = 1e$^{-4}$ and $T_{tol}$ = \SI{1}{\kelvin}. All cases are simulated over 1,000 CFD time steps with a maximum Courant number of 0.5.
 
\begin{figure}[H]
\centering
{\includegraphics[width=120mm]{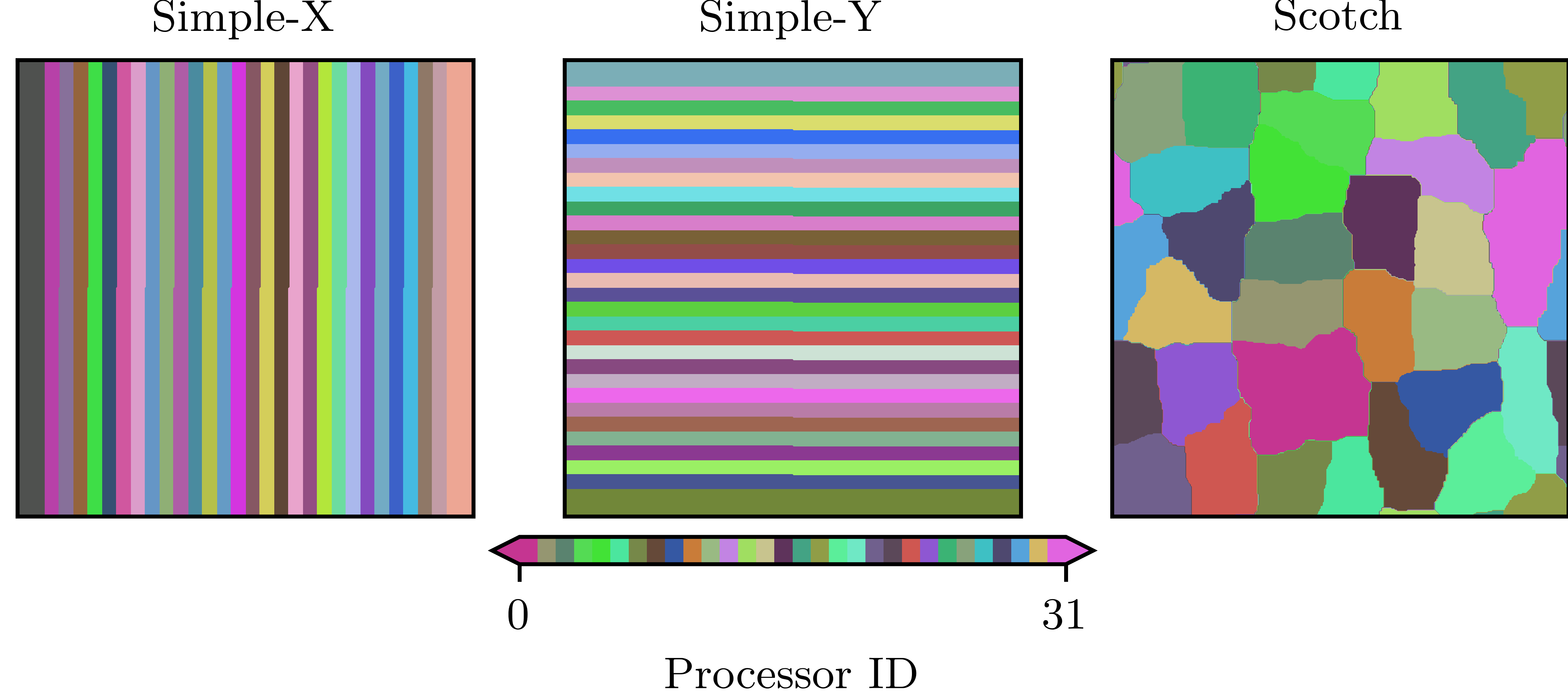}}
\caption{Decomposition of the test case to 32 processes using 1) Simple decomposition in X direction (left), 2) Simple decomposition in Y direction (middle) and 3) Scotch decomposition (right). While Simple-X attempts to reduce load imbalance, Simple-Y promotes it. Scotch decomposition by design attempts to minimize the number of process boundaries. {Data, plotting scripts and figure are available under the CC-BY license \cite{data}.}}
\label{fig:decomposition}
\end{figure}

The reactivity (and the computational load) of the case is higher within the shear layer between fuel and oxidizer across X direction. Therefore, this case can be decomposed using different approaches to have high or low load imbalance. Different decomposition approaches tested here are illustrated in Figure \ref{fig:decomposition}. While a simple decomposition in X direction (left) would ensure minimal imbalance since the decomposition is orthogonal to the shear layer, the same decomposition in Y direction (middle) would result in higher imbalance since now only the processes assigned to/around the shear layer will have high load, creating a bottleneck. Finally, the Scotch decomposition (right) is also investigated since in many CFD applications automated and optimized decomposition methods are used to ensure uniform cell counts and minimize the process boundaries. In addition, since the majority of the cells in the configuration are initially pure oxidizer (and identical), the reference mapping method can be used to significantly reduce the computational load in the processes assigned to the oxidizer part,  increasing the balancing performance. Therefore, the reactive shear layer is a good test case to investigate: i) the effect of load imbalance on $\chi_{su}$ and ii) the additional performance that can be gained from using the reference mapping method together with the load balancer. The details of the parameters investigated in this case (decomposition, number of processors and balancing model) are given in Table \ref{table:shearlayer}.

\begin{figure}[H]
\centering
{\includegraphics[width=0.6\columnwidth]{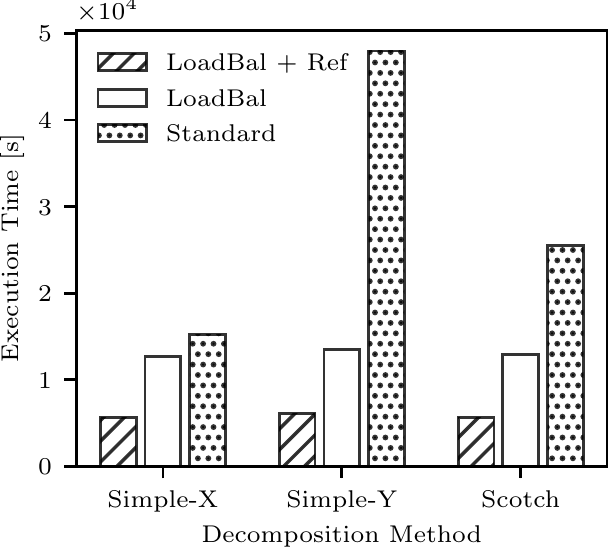}}
\caption{A bar chart showing the execution times for Simple-X, Simple-Y and Scotch decomposition methods for \nproc = 32. Increase in speed-up with increased imbalance can be observed particularly from Simple-Y decomposition. {Data, plotting scripts and figure are available under the CC-BY license \cite{data}.}}
\label{fig:shear_64}
\end{figure}

\begin{center}
\captionof{table}{The difference decomposition methods, number of processors, and balancing models investigated in reactive shear layer simulations.} \label{table:shearlayer}
\resizebox{0.8\columnwidth}{!}{%
\begin{tabular}{@{}c|c@{}}
\textbf{Decomposition} & Simple-X, Simple-Y, Scotch \\
\hline
\textbf{Number of processors} & 32, 64, 128, 256 \\
\hline
\textbf{Model} & \begin{tabular}[c]{@{}c@{}}Standard,\\  Load Balanced,\\ Load Balanced + Reference Mapping\end{tabular}
\end{tabular}}
\end{center}
The results obtained from the 3 different decomposition methods for \nproc = 32 are presented in Figure \ref{fig:shear_64}.  It can be seen that the total execution times are significantly reduced when the load balancing is utilized. While the domain decomposition plays an important role on the performance of the standard case with no balancing, for load balanced models the execution time is not dependent on the decomposition strategy. For instance, while the Simple-Y decomposition has the highest execution time when using the standard model due to load imbalance caused by the decomposition, the execution times of load balanced models are similar to other decomposition methods. Finally, it is important to mention that while the load balancer alone already provides a substantial performance increase, utilizing the reference mapping model further increases the performance of the balancer by reducing the mean load of the simulation and creating further imbalance to increase the balancing performance.

\begin{figure}[H]
\centering
{\includegraphics[width=0.5\columnwidth]{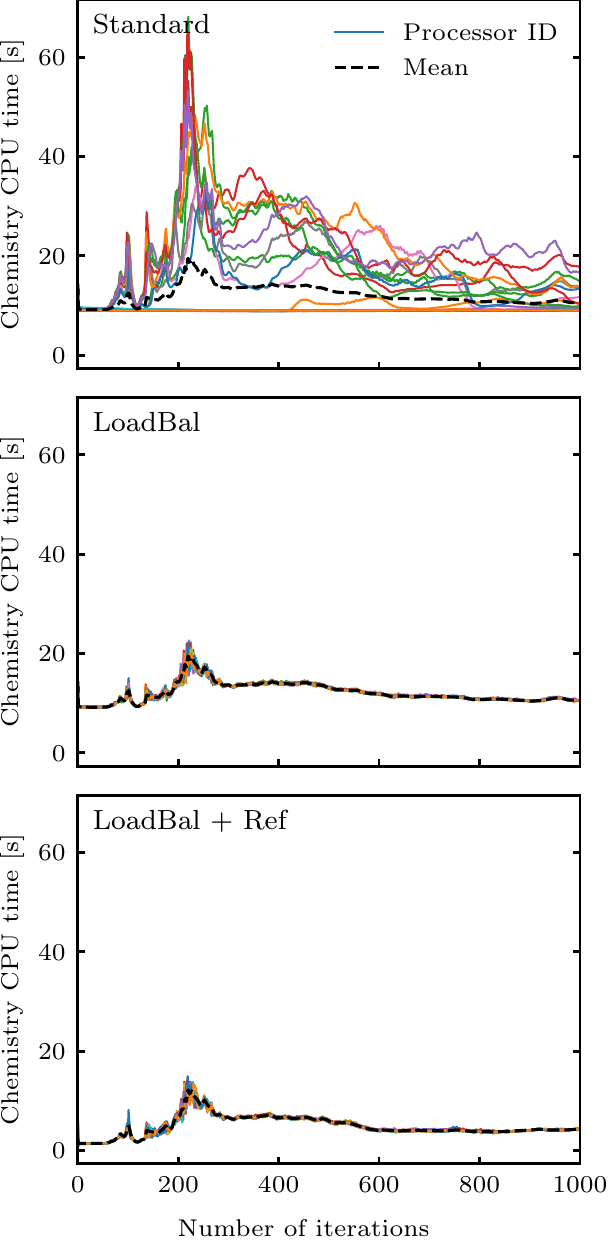}}
\caption{Chemistry solution CPU time for each process and the corresponding arithmetic mean in each configuration for the reactive shear layer problem. It is noted that the deviation from the mean value is higher for the standard model, which causes computational performance issues. {Data, plotting scripts and figure are available under the CC-BY license \cite{data}.}}
\label{fig:rankbased}
\end{figure}

To further demonstrate the effect of load balancer on reactive simulations, the load, i.e.~CPU time for solving the chemistry ($\tau_{\mathrm{cpu}}^k$) on each process along with its arithmetic mean across processes is presented in Figure \ref{fig:rankbased} for the case with \nproc = 32 and Scotch decomposition. It is important to note that the profiles given here only show the time spent on solving the chemistry and do not include the time spent on balancing, which is not significant compared to $\tau_{\mathrm{cpu}}^k$ as shown in Section \ref{section:benchmark}. It can be seen that for the non-balanced case (bottom), the deviation of $\tau_{\mathrm{cpu}}^k$ with respect to its mean value is very high. In this scenario, the most computationally loaded process becomes the bottleneck, while the other processes are waiting for that process to finish the chemistry computation. For load balanced cases (middle and bottom), there is a very small deviation in $\tau_{\mathrm{cpu}}^k$ between processes with respect to the arithmetic mean, which eliminates the bottleneck caused by the unbalanced chemistry load and reduces the simulation time. 

\begin{figure}[H]
\centering
{\includegraphics[width=0.6\columnwidth]{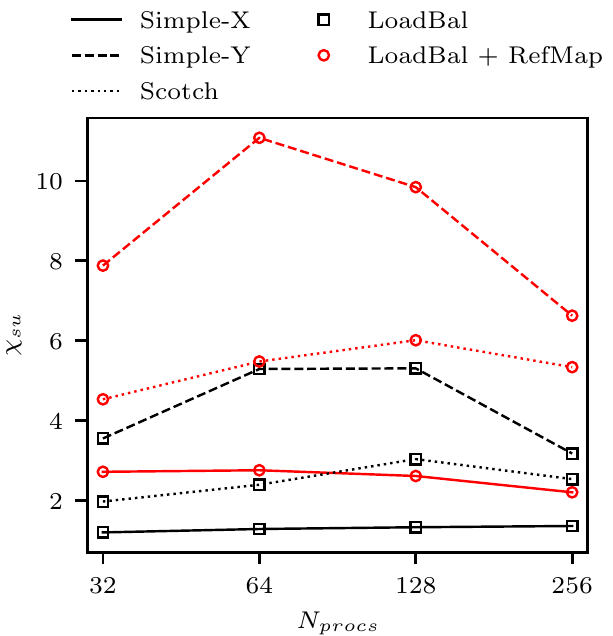}}
\caption{Speedup factor for all the cases described in Table \ref{table:shearlayer}. We note that the highest speedup is achieved with Simple-Y decomposition with reference mapping utilized, due to imbalance created by decomposition and balancing gain obtained by reference mapping, respectively. {Data, plotting scripts and figure are available under the CC-BY license \cite{data}.}}
\label{fig:shearlayer_speedup}
\end{figure}

Finally, the $\chi_{su}$ for all cases described in Table \ref{table:shearlayer} with respect to the standard model are reported in Fig.~\ref{fig:shearlayer_speedup}. For the load balanced model, the $\chi_{su}$ is small for Simple-X decomposition due to already balanced load distribution. However, for the other 2 decomposition methods a $\chi_{su}$ around 2 to 5 can be observed. When the reference mapping is also utilized, we observe that $\chi_{su}$ increases even further. While for Scotch decomposition $\chi_{su}\approx$ 4 to 6 is achieved, for Simple-Y decomposition we can observe $\chi_{su}\approx$ 8 to 12.

\subsection{Three dimensional reacting diesel spray}
\label{section:sprayA}
\begin{figure}[H]
\centering
{\includegraphics[width=0.9\columnwidth]{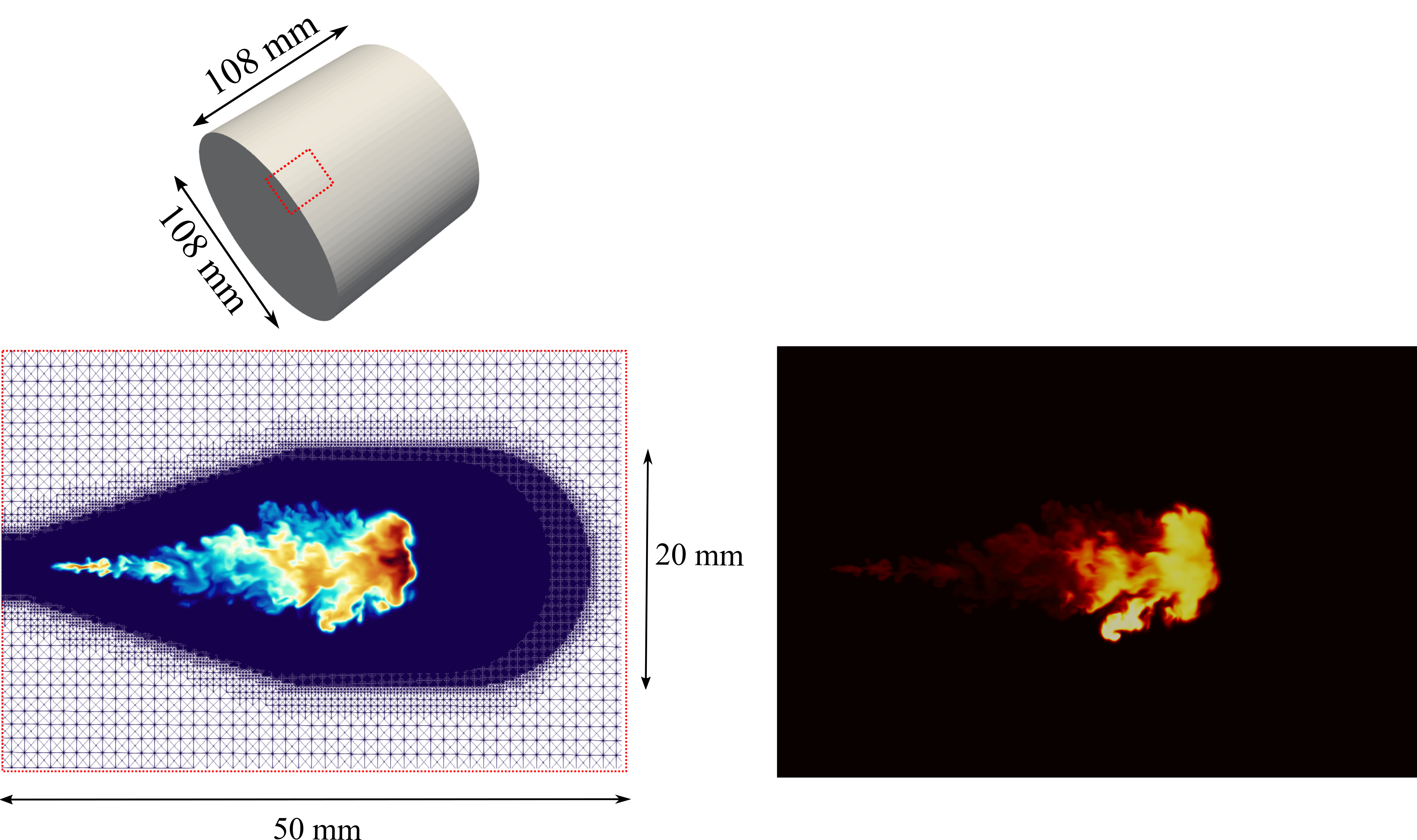}}
\caption{Schematic of the test case and the computational grid used in spray simulations based on our earlier work \cite{Kahila2019,Kahila2019a,Tekgul2020} (left) and an instantaneous representation of the temperature field during the simulation (right). In left figure, the mixture fraction (\textit{Z}) field is used for visualization, which gives information about the fuel mass fraction and spray vapor.}
\label{fig:mesh}
\end{figure}

To further demonstrate the performance of the implemented model, a larger and more challenging case is simulated in this subsection. A 3D reactive spray LES with liquid fuel injection, featuring the Engine Combustion Network Spray A condition \cite{cite_ECN} is utilized. The details of this simulation configuration can be found in our earlier work \cite{Kahila2019,Tekgul2020}. A summary of the case setup is given in Table \ref{table:ecncase}. A cylindrical constant volume domain with dimensions 108x108 mm and a \SI{1}{\milli\meter} base mesh along with a uniform mesh refinement region of \SI{125}{\micro\meter} enclosing the spray is used.  A total of $\approx$4.5 M cells are decomposed into 256 processes with Scotch decomposition, resulting with $\approx$17,000 cells per process. A constant CFD time step of 2e$^{-7}\:\mathrm{s}$ is utilized. The same chemical mechanism used in the reactive shear layer case was also used here. {The reference cell would be chosen as a non-reactive ambient cell with $Z_{tol}$ =1e$^{-4}$ 1e$^{-4}$ and $T_{tol}$ = \SI{1}{\kelvin}.}  Due to the poor performance of the standard model and our computational limitations, all simulations are started from a time instance prior to ignition, and simulated for 500 timesteps instead of the full simulation. A schematic demonstrating a portion of the computational domain is presented in Figure \ref{fig:mesh}. 

\begin{center}
\centering
\captionof{table}{Spray case setup details. See \cite{cite_ECN} for further details.}
\label{table:ecncase}
\begin{tabular}{ll|l}
                  & \multicolumn{1}{l}{\begin{tabular}[c]{@{}l@{}}\\ \end{tabular}} & \textbf{ECN Spray A}                                                                        \\ 
\cline{2-3}
\multirow{5}{*}{} & Injected fuel                                                            & \textit{n}-$\ce{C12H26} $                                                                   \\
                  & Nominal nozzle diameter, D                                      & \SI{90}{\micro\meter}                      \\
                  & Injection pressure                                              &                                        \SI{150}{\mega\pascal}  \\
\cline{2-3}
\multirow{6}{*}{} & Temperature                                                     & \SI{900}{\kelvin}                                                                                        \\
                  & \ce{O2} molar fraction                                                   & 0.15                                                                                        \\
\end{tabular}
\end{center}

\begin{figure}[H]
\centering
{\includegraphics[width=0.5\columnwidth]{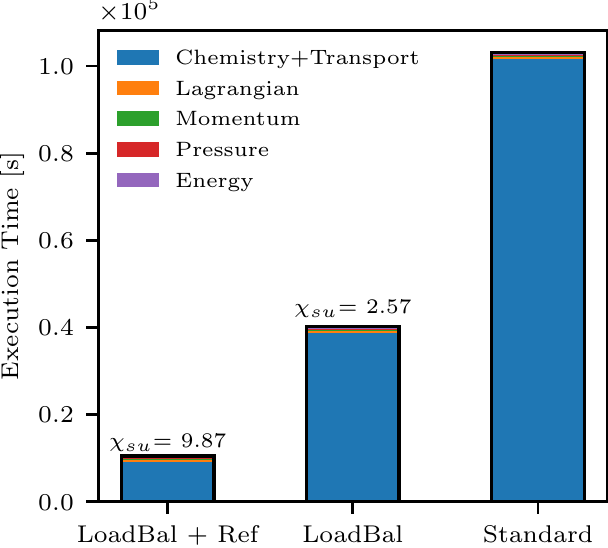}}
\caption{Speedup for different models. A speed-up by a factor of 2.57 and 9.87 obtained with load balancing and load balancing + reference mapping, respectively. {Data, plotting scripts and figure are available under the CC-BY license \cite{data}.}}
\label{fig:spray_speedup}
\end{figure}

Figure \ref{fig:spray_speedup} shows the total execution times for the 3 investigated cases: standard, load balancing and load balancing with reference mapping. The computational cost of different solver operations are also timed and presented in the figure. It can be observed that the computational cost of flow equations (momentum, pressure, energy, and lagrangian) takes up a very small fraction of the total execution time. In contrast, the chemistry and species transport equations take up about 90 to 98 \% of the total execution time. A speedup $\chi_{su}$ = 2.57 is obtained compared to the standard model when load balancing is utilized. When the reference mapping is also utilized, the $\chi_{su}$ increases to 9.87. The gain obtained by using the reference mapping model is higher in this case compared to the reactive shear layer case presented in Section \ref{subsec:shearlayer} due to the transient nature of the spray injection. The chemistry is solved within the spray region which takes up a small percentage of the total cells at the early stages of the simulation. Utilizing the reference mapping provides further speedup due to the large number of idle processes outside of the spray region.

\section{Conclusions and future work}
\label{section:conclusions}

A dynamic load balancing and a reference mapping model named \texttt{DLBFoam} to speed up the parallel reactive CFD simulations in OpenFOAM is presented and investigated. Dynamic load balancing utilizes MPI routines to send chemistry problems from processes with higher computational load to ones with less load, to avoid computational bottlenecks in parallel simulations. In addition, the reference mapping acts as a filter, where the chemical source terms of the cells satisfying a given criteria are copied from a reference solution, instead of direct integration.

A thorough performance analysis of the model has been presented, showing the balancing efficiency to scale up to a high number of processes and number of problems per process. Furthermore, the model has been utilized on two reactive cases: a 2D reactive shear layer and a 3D diesel spray combustion case. Significant speed-up up to a factor of 10 to 12 has been achieved and demonstrated when load balancing and reference mapping models are used together. It is worth noting that the reported speed-up results are for the particular cases investigated, as well as the ODE solver tolerances chosen and the chemical mechanism used. Changing ODE convergence tolerances and using larger/smaller chemical mechanisms may result in less or more favorable results in terms of load balancing performance.

In addition to the improvements we have introduced in this study, there are still certain aspects that can be modified to further increase the performance of the load balancing model and extend its compatibility. Two potential improvements are following:  
\begin{itemize}
    \item \textit{Improving performance and accuracy of the ODE solvers}: Our experience on state-of-the-art ODE solution algorithms including analytical Jacobian \cite{Niemeyer2016} and problem-type tailored linear algebra \cite{Kahila2019,Imren2016} is noted to significantly decrease ODE integrator iterations, enhance its robustness and even further increase the load-balancing efficiency. 
    \item \textit{Extension to in-situ tabulation and reduction models}: Presently, the implementation is not compatible with in-situ adaptive reduction and tabulation model TDAC available in OpenFOAM. Extending the load balancer to these models will both increase the load balancing performance and mitigate the performance issues still present in TDAC due to its processor-based formulation.
\end{itemize}
Including these features to our open-source library will be investigated in the near future.

\section{Conflict of Interest}
We wish to confirm that there are no known conflicts of interest associated with this publication and there has been no significant financial support for this work that could have influenced its outcome.

\section*{Acknowledgements}
\label{}
The present study has been financially supported by the Academy of Finland (grant number 318024). The computational resources for this study were provided by CSC - Finnish IT Center for Science. The first author has been financially supported by the Merenkulun S\"{a}\"{a}ti\"{o}. 
\appendix

\setcounter{figure}{0}

\section{Reference mapping model validation}
\label{appendixA}

\begin{figure}[H]
\centering
{\includegraphics[width=0.6\columnwidth]{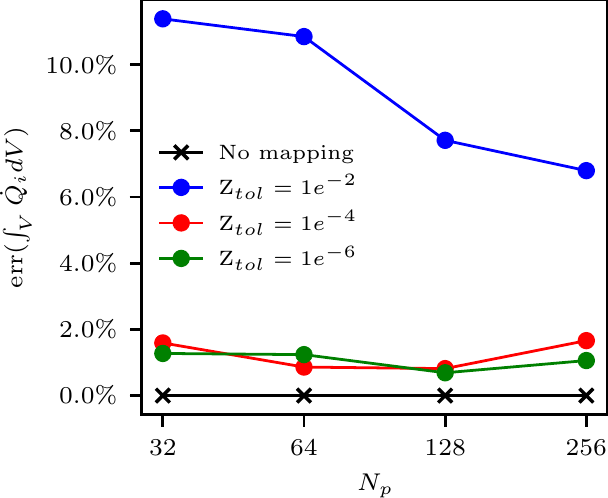}}
\caption{\label{fig:validation1} The absolute percentage error of volume integrated heat release rate averaged over all time steps for different process counts. {Data, plotting scripts and figure are available under the CC-BY license \cite{data}.}}
\end{figure}

The purpose of this appendix is to quantify the error introduced by the reference mapping model. The error introduced by using the reference mapping model is illustrated in Figure~\ref{fig:validation1}.  The figure shows the mean absolute percentage error of the volume integral of the heat release rate over all time instances relative to the results obtained with the standard chemistry model in OpenFOAM. The results are obtained from the reactive shear layer case in Section \ref{subsec:shearlayer} with different process counts. {$Z_{tol}$ = 1e$^{-2}$, 1e$^{-4}$, 1e$^{-6}$} and $T_{tol}$ = \SI{1}{\kelvin} values are used as model tolerance. When the load balancing algorithm is used without mapping, identical results to the standard model are obtained for all process counts as expected. When the reference mapping is used, {initially a large deviation around 10\% is observed for $Z_{tol}$ = 1e$^{-2}$. With tighter tolerances, we observe this error to drop below 2 \% and converge between $Z_{tol}$ = 1e$^{-4}$ to 1e$^{-6}$. This observation is consistent with our earlier studies, in which a $Z_{tol}$ = 1e$^{-4}$ value is successfully used to model various spray combustion phenomenon \cite{Kahila2019,Kahila2019a,Tekgul2020}.} The deviation from the standard model depends on the process count due to the process-local nature of the reference mapping. {A video animation demonstrating how the reference mapping operates during the simulation is provided in the supplementary materials of the paper.}   

\begin{figure}[H]
\centering
{\includegraphics[width=0.6\columnwidth]{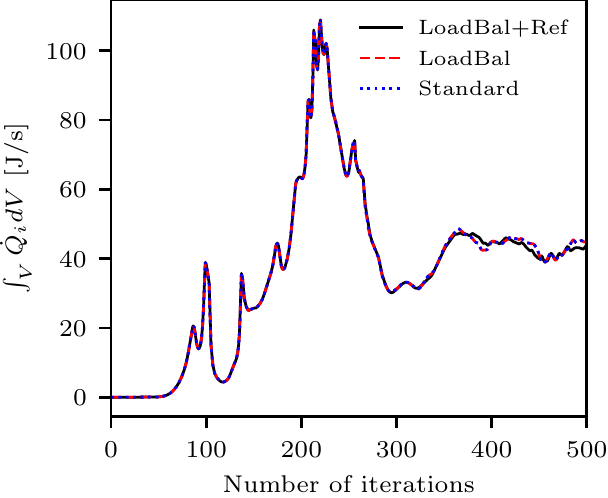}}
\caption{\label{fig:validation2} Volume integrated heat release rate for \nproc = 64, using load balanced + reference mapped ({with $Z_{tol}$ = 1e$^{-4}$}), load balanced and standard models. {Data, plotting scripts and figure are available under the CC-BY license \cite{data}.} }
\end{figure}

The evolution of the volume integrated heat release rate is given in Figure \ref{fig:validation2}. As seen, the reference mapping shows a very good match with the non-mapped models for the presented simulation data. It is important to note that the error due to the reference mapping may accumulate over time depending on the model tolerance values used. Although the performance gain when using the reference mapping is significant compared to the introduced error, for applications that require very high accuracy, a careful model tolerance analysis is recommended.

\setcounter{figure}{0}

\section{Computational overhead}
\label{appendixB}
 {This appendix quantifies the computational overhead introduced by the load balancing algorithm. As described in Section \ref{subsec:balancing}, the dynamic load balancing algorithm consists of 4 different stages: 1) preprocessing stage, 2) balancing stage, 3) solution stage, and 4) update stage. Among those, stages 1,2 and 4 can be considered in computational overhead context. Although we have noted the overhead to be insignificant compared to the solution time, here we quantify the overhead for the three dimensional reacting diesel spray benchmark case we utilized at the end of the paper.}

\begin{figure}[H]
\centering
{\includegraphics[width=0.5\columnwidth]{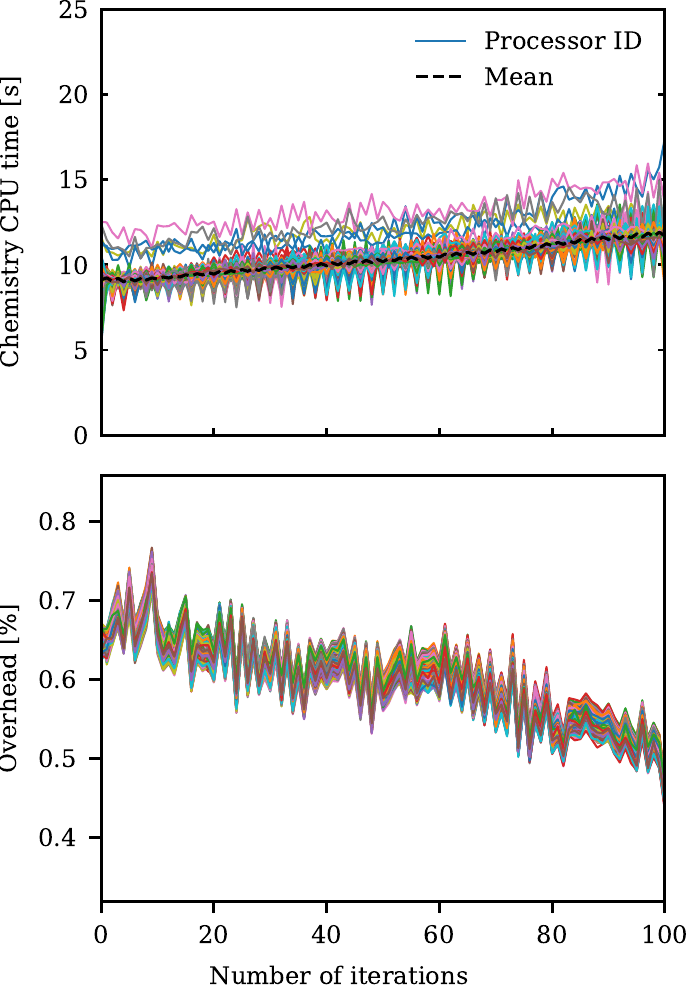}}
\caption{Chemistry solution CPU time and corresponding computational overhead percentage related to dynamic load balancing for each process. {The solid colored lines represent the CPU time of each processor, while the black dashed line represents their arithmetic mean.} The computational overhead accounts for less than 1\% of the chemistry solution time. {Data, plotting scripts and figure are available under the CC-BY license \cite{data}.}}
\label{fig:overhead}
\end{figure}

{Figure \ref{fig:overhead} shows the CPU time spent on solving the chemistry and the percentage of computational overhead with respect to the chemistry CPU time for 100 CFD iterations. {Following the case setup described in Section \ref{section:sprayA}, the domain is decomposed into 256 processors.} It should be noted that the idling time associated with the explicit barrier at the end of the stage 4 is not counted towards computational overhead. It can be seen that the computational overhead associated with load balancing is less than 1\% of the total chemistry solution CPU time. While the overhead may increase by increasing the cells per process or the size of the utilized chemical mechanism, we note that it is insignificant for cases and mechanisms of relevant size to reactive CFD applications.} {We note that Figure \ref{fig:overhead} shows outlier processors, i.e., processors with slightly higher load than the mean value after balancing. Based on our experience, this behavior is due to 1) transient nature of the process-based load value due to local chemistry stiffness, and 2) possible hardware and parallel communication issues that are briefly mentioned in Section \ref{section:benchmark}. These aspects exist in any reactive CFD simulation running on any hardware architecture, and are not caused by the balancing algorithm.}

\section{Supplementary material}

All the analysis in this paper is performed using the \texttt{DLBFoam}. \texttt{DLBFoam} is open-source and publicly available at \url{https://github.com/blttkgl/DLBFoam}. The repository contains instructions for compilation and tutorials. {The simulation data, plotting scripts, and the figures featured in this paper are available as supplementary material under the CC-BY license \cite{data}.}
\bibliography{library.bib} 
\bibliographystyle{elsarticle-num.bst}

\end{document}